# X-ray evolution in the ROSAT Brightest Cluster Sample

H. Ebeling[1], S.W. Allen[1], C.S. Crawford[1], A.C. Edge[1], A.C. Fabian[1], H. Böhringer[2], W. Voges[2], J.P. Huchra[3]

[1] Institute of Astronomy, Madingley Road, Cambridge CB3 0HA, UK
[2] Max-Planck-Institut für extraterrestrische Physik, Giessenbachstrasse, 85740 Garching, Germany
[3] Harvard-Smithsonian Center for Astrophysics, 60 Garden Street, Cambridge, MA 02138, USA

**Abstract.** We present the X-ray luminosity function (XLF) of the ROSAT Brightest Cluster Sample (BCS), an X-ray selected, flux limited sample of 172 clusters of galaxies at $z \leq 0.3$ compiled from ROSAT All-Sky Survey data. While the bulk of the BCS consists of Abell clusters, the sample also contains Zwicky clusters and purely X-ray selected systems. The BCS-XLF represents the best determination of the local X-ray luminosity function for galaxy clusters and thus provides an important reference for evolutionary studies. For the BCS itself, we find no convincing evidence for cluster evolution within a redshift of $z = 0.3$. This result is not in conflict with the findings of the EMSS study on cluster evolution.

## 1. Introduction

Clusters of galaxies represent ideal targets for studies on the formation of gravitationally bound structure on very large scales. Ideally, cosmologists want to know the cluster mass function at various epochs in order to test the predictions of cosmological models of structure formation. Unfortunately, with present-day observational techniques the total mass of a cluster is not a directly observable quantity. The cluster X-ray luminosity function (XLF), however, is closely related and has the advantage that it can be established comparatively straightforwardly.

The existence of evolution in the cluster X-ray luminosity function, i.e. in the space density of clusters of a given X-ray luminosity with redshift, has been a hotly debated issue over the past few years. Edge et al. (1990) found strong evidence for evolution in a flux limited sample of 55 clusters within a redshift of 0.18 in the sense that X-ray luminous clusters were less common in the past than they are today. At higher redshifts, the same effect of negative evolution was observed by Gioia et al. (1990) and Henry et al. (1992) in a sample of similar size compiled from EINSTEIN Medium Sensitivity Survey (EMSS) data. Despite their apparent similarity, the findings from these studies are nonetheless difficult to reconcile. Both imply such strong evolution that it is difficult to provide a self-consistent model to match them.

Using an X-ray flux-limited sample of more than 250 ACO clusters selected from ROSAT All-Sky Survey (Voges 1992, Trümper 1993) data, Ebeling et al. (1994) resolved this conflict by showing that the apparent strong evolution found by Edge and co-workers at high X-ray fluxes was the result of a significant dearth of high-luminosity clusters in the redshift range from 0.1 to 0.15. The XLF for the much larger sample of ACO clusters of Ebeling et al. showed only very mild negative evolution out to redshifts of $z \sim 0.2$. However, since their sample contains only ACO clusters, it combines optical and X-ray selection criteria and is thus not in all respects a fair sample of clusters in general. The need for a representative sample of clusters at low to moderate redshifts is all the more urgent as there is evidence that, at much higher redshifts ($z \leq 0.9$), clusters may evolve very rapidly (Castander et al. 1995). [The latter claim is, however, based on a sample of only 13 clusters, and more work is required to confirm or discard this result (see Rosati, Jones et al., and Burke et al., these proceedings).]

## 2. The ROSAT Brightest Cluster Sample

The ROSAT Brightest Cluster Sample (BCS) was conceived as a flux limited sample of X-ray selected galaxy clusters detected during the ROSAT All-Sky Survey (RASS) in the northern hemisphere ($\delta \geq 0°$), and at high Galactic latitude ($|b_{II}| \geq 20°$). Although ACO clusters constitute the bulk of the BCS, it also contains Zwicky clusters and systems selected on the grounds of their X-ray extent alone. Extensive optical follow-up work (Allen et al. 1992, Crawford et al. 1995) accompanied the compilation of the sample in the X-ray.

In a nutshell, the compilation proceeded as follows: starting from a list of all sources detected by the Standard Analysis Software System (SASS, Voges et al. 1992) in the first processing of the RASS data, we selected all sources with SASS count rates in excess of $0.1$ s$^{-1}$ in the extragalactic part of the northern hemisphere. This source list was cross-correlated against the Abell and Zwicky cluster catalogues. In addition to the coincidences found in the cross-correlation, we also included all sources with a value for the SASS extent parameter of at least 35 arcsec to our tentative sample. In order to improve upon the fluxes re-turned by the SASS (which, by design, is a point-source detection algorithm) we then reprocessed the RASS photon maps in $2 \times 2$ deg$^2$ fields around our cluster candidates using VTP (Ebeling & Wiedenmann 1993), an algorithm optimized for the detection of extended and irregularly shaped emission. The cross-correlation was then repeated with the VTP source list and sources classified as extended by VTP were added to the sample. All cluster candidates were then scrutinized in the X-ray and in the optical, and non-cluster sources were removed. By and large, the cluster selection follows very much the procedure employed in the compilation of the mentioned ACO cluster sample [see Ebeling et al. (1995) for details].

Figure 1 shows the luminosity-redshift distribution of the BCS; the flux limit in the 0.1 to 2.4 keV band (indicated by the dotted line in Fig. 1) is $5.5 \times 10^{-12}$ erg cm$^{-2}$ s$^{-1}$. $H_0 = 50$ km s$^{-1}$ Mpc$^{-1}$ and $q_0 = 0.5$ are used throughout.

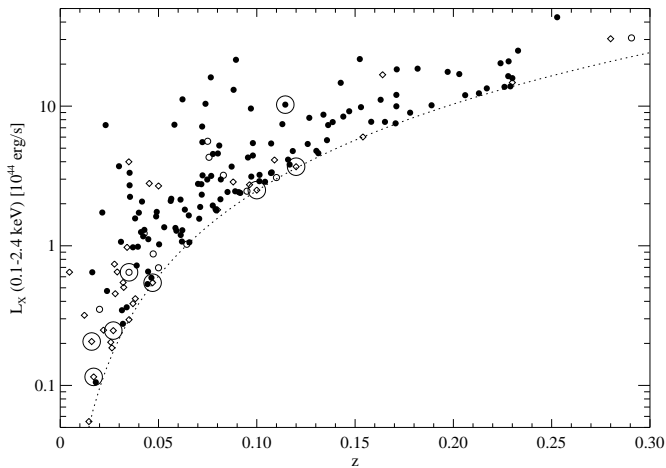

**Fig. 1.** The distribution of the 172 clusters of the BCS as a function of redshift. Abell clusters are plotted as filled circles, Zwicky clusters as open circles. Clusters selected on the grounds of their SASS or VTP extent are shown as open diamonds. Clusters detected serendipitously by VTP during the reprocessing appear encircled.

## 3. The XLF for the BCS

As can be seen from Fig 1, the BCS contains a number of clusters missed by the SASS but detected serendipitously by VTP in the $2 \times 2$ deg$^2$ fields around the initial SASS cluster candidates. Since the total area covered by these fields amounts to only 16 per cent of our study area (i.e. the northern extragalactic sky), the BCS is actually not complete at the quoted flux limit. However, for the XLF we can correct for this incompleteness by weighting the serendipitous detections accordingly.

Figure 2 shows the XLF for the BCS after correction for incompleteness. The XLF is well described by a Schechter function

$$\frac{dn}{dL_X}(L_X) = A \, \exp(-L_X/L_X^\star) \, L_X^{-\alpha} \qquad \text{with}$$

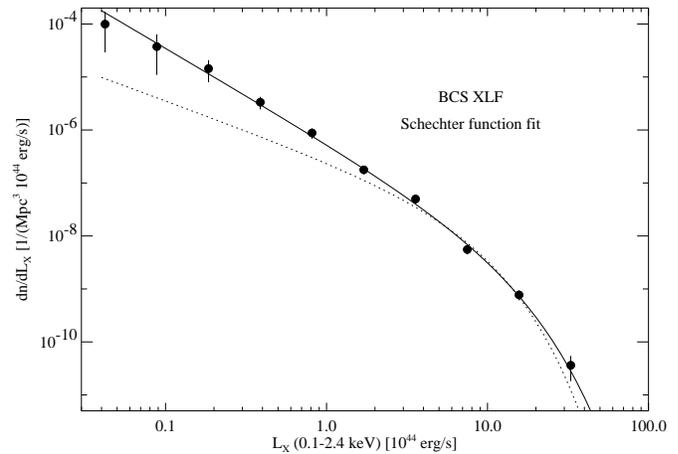

**Fig. 2.** The X-ray luminosity function for the BCS. The dotted line shows the Schechter function fit to the XLF for the ACO cluster sample of Ebeling et al. (1994).

$$A = (5.72 \pm 0.95) \, 10^{-7} \, \tfrac{1}{\text{Mpc}^3} \, (10^{44} \text{ erg/s})^{\alpha-1},$$
$$L_X^\star = (8.92 \pm 1.66) \, 10^{44} \text{ erg/s}, \quad \alpha = 1.78 \pm 0.09.$$

A first analysis of the BCS-XLF in different redshift shells confirms the mild (or no) evolution result of Ebeling et al. (1994). However, whereas the ACO cluster sample of Ebeling et al. (1994) is limited to redshifts below 0.2, the BCS extends out to $z = 0.3$ and thus has substantial overlap with the redshift domain of the EMSS sample. The lack of significant negative evolution in the BCS is not in conflict with the findings of Gioia, Henry and coworkers. The EMSS study finds a significant steepening in the power law approximation of the XLF only when clusters at redshifts between 0.3 and 0.6 are considered; within $z \leq 0.3$ their data are perfectly consistent with a no-evolution model.